\documentclass[12pt]{amsart}
\usepackage[english]{babel}
\usepackage{amsmath,amssymb}
\usepackage{graphicx}
\usepackage[top=3cm,left=3cm,right=3cm,bottom=3cm]{geometry}
\usepackage{cite}
\begin{document}

 \title[Statistical analysis of the Ibex35 index]{Scaling, stability and distribution of the high-frequency returns
of the \textsc{Ibex35} index}
 \author{Pablo Su\'arez-Garc\'{\i}a}
 \email{pasuarez@fis.ucm.es}
 \author{David G\'omez-Ullate}
 \email{dgullate@fis.ucm.es}
 \address{Departamento de F\'isica Te\'orica II, Universidad Complutense, 28040
Madrid, Spain.}


\begin{abstract}
In this paper we perform a statistical analysis of the high-frequency returns of the \textsc{Ibex35} Madrid stock exchange index. We find that its probability distribution seems to be stable over different time scales, a stylized fact  observed in  many different financial time series. However, an in-depth analysis of the data using maximum likelihood estimation and different goodness-of-fit tests rejects the L\'evy-stable law as a plausible underlying probabilistic model. The analysis shows that the Normal Inverse Gaussian distribution provides an overall fit for the data better than any of the other subclasses of the family of the Generalized Hyperbolic distributions and  certainly much better than the L\'evy-stable laws. Furthermore, the right (resp.\ left) tail of the distribution seems to follow a power-law with exponent \mbox{$\alpha \approx 4.60$} (resp.\  \mbox{$\alpha \approx 4.28$}). Finally, we present evidence that the observed stability is due to temporal correlations or non-stationarities of the data.\\[8pt]

%In this paper we perform a statistical analysis of the high-frequency returns of the \textsc{Ibex35} Madrid stock exchange index. We find that the probability distribution function seems to be stable over different time scales, a stylized fact also observed in  many different financial time series. However, an in-depth analysis of the data using maximum likelihood estimation and different goodness-of-fit tests rejects the L\'evy-stable law as a plausible underlying probabilistic model.  Even though the right (resp.\ left) tail of the distribution seem to follow a power-law with exponent \mbox{$\alpha \approx 4.60$} (resp.\  \mbox{$\alpha \approx 4.28$}), the analysis shows that the Normal Inverse Gaussian distribution provides an overall fit for the data better than any of the other members of the family of the Generalized Hyperbolic distributions and  certainly much better than the L\'evy-stable laws. Finally, we present evidence that the observed stability is due to temporal correlations or non-stationarities of the data.\\[8pt]
\end{abstract}
\maketitle



% \linenumbers

\section{Introduction}
\label{Introduction}

The marginal distribution of returns of financial assets have been placed under scrutiny since the times of Bachelier \cite{bachelier1900theorie}, and the idea of treating log-returns as independent identically distributed Gaussian random variables  lies in the core of the most well-known and celebrated financial models  \cite{black1973pricing, merton1973theory} and so it is crucial for derivative pricing and risk management. And even though this idea works fine as a first approximation, it is well documented that  empirical financial data drawn from very different markets, time periods and instruments do not fit the Gaussian model \cite{mandelbrot1963variation, fama1965behavior, cont1997scaling, pagan1996econometrics}. The empirical distributions of log-returns present  tails heavier than Gaussian as well as many other non-trivial statistical properties collectivelly know as \emph{stylized facts} \cite{cont2001empirical} that place the Gaussian hypothesis in jeopardy and point towards a possible universal behavior of the underlying processes.

One of the most celebrated of these \emph{stylized facts} is the \emph{scaling symmetry} or \emph{stability} of the distribution of log-returns, \emph{i.e.}\ its invariance under aggregation up to rescaling. For  independent identically distributed random variables the Gaussian law is the only distribution with finite second moment that has this property, and that is why the central limit theorem singles it out as the limiting distribution of rescaled sums of \emph{i.i.d.}\ random variables with finite variance \cite{breiman1968probability}. To observe stability in distributions other than Gaussian the requirement of a finite second moment has to be dropped. This seminal idea was first expounded in finance by Mandelbrot \cite{mandelbrot1963variation} who proposed the family of L\'evy-stable laws as an alternative to the Gaussian model of log-returns. One  feature of these probability distributions is the divergence of their second moment caused by the power-law behavior of its tails with characteristic exponent $\alpha<2$. Considering a financial market as a complex system of interacting agents in which prices are the outcome of many independent individual decisions, according to the generalized central limit theorem its limiting distribution should be a member of the L\'evy-stable family ---\,of which the Gaussian distribution is just a special case\,--- with the normalization constant depending on the tail index of the power-law \cite{breiman1968probability}.

Therefore, it seems natural to investigate the tails of the distribution of log-returns in order to shed some light on its stability properties. However, this is a moot point: although some authors \cite{mandelbrot1963variation, fama1965behavior} have reported power-law behavior with $\alpha<2$,  others have reported distributions with power-law tails far away from the stability regime \cite{gopikrishnan1999scaling, lux1996stable, lux2001limiting, plerou1999scaling, farmer1999physicists}. It could be argued that there is an endemic arbitrariness of the least-square regression used to study the power-law behaviour in empirical data, but that problem could be overcome replacing it with maximum likelihood estimation together with goodness-of-fit techniques  \cite{clauset2008power}. This fact notwithstanding, it is also well known that an estimated tail index above two is not an evidence against stability:  it could well have been produced by a  stable distribution with $\alpha$ as low as 1.65 with the situation getting worse as we approach the $\alpha=2$ limit \cite{dumouchel1983estimating, mcculloch1997measuring}. And to round this off, it is not only difficult to discriminate between different power-laws or even between stability or the lack of it;  the sole task of distinguishing a power-law from a stretched exponential is still subject to debate \cite{malevergne2005empirical}, since certain empirical distributions of log-returns seem to decay asymptotically slower than any power-law \cite{gourieroux1998truncated, laherrere1998stretched, cont1997scaling}.

Among the distributions with tails lighter than power-laws, a family that has been used with success to model log-returns are the \emph{Generalized Hiperbolic laws}. The embryo of this family of distributions is the \emph{Hyperbolic distribution}, first proposed by Ralph Alger Bagnold \cite{bagnold1941physics} to model the size distribution of  the wind-blown sand. Later, Barndorff-Nielsen ---\,still with the problem of the distribution of particle size in mind\,--- generalized it to the family of \emph{Generalized Hyperbolic} (\emph{GH})  distributions  \cite{barndorff1977exponentially}, of which the hyperbolic distribution is  a special case. Different subclasses of this family have been since then proposed as alternatives to both Gaussian and L\'evy-stable laws as statistical models of financial returns, namely, the \emph{Skewed Student's t} distribution  \cite{Praetz1972, blattberg1974comparison}, the \emph{Hyperbolic distribution} by Eberlein and Keller  \cite{eberlein1995hyperbolic, kuchler1999stock}, the \emph{Variance-Gamma} of Madan and Seneta   \cite{madan1990variance} and the \emph{Normal Inverse Gaussian} (\emph{NIG}) by Barndorff-Nielsen  \cite{barndorff1995normal}. One of the most appealing properties of this family is its tail behavior, which is a power-law modulated by an exponential. These lighter tails seem well suited to fit the empirical distributions of log-returns as the studies cited above show, since the data seem to have tails heavier than Gaussian but still lighter than the L\'evy-stable laws. 

As can be inferred, the question of the true nature of the  distribution of log-returns (or even its tail behavior) and the origin of its apparent stability is far from being settled, and therefore, the study of diverse financial time series drawn from different instruments, markets and time periods is pertinent in order to shed some light on this issue. In this paper we will carry out a thorough study of the distributional properties of the high-frequency log-returns of the index \textsc{Ibex35} from the Madrid Stock Exchange. After reviewing the basic properties of both the L\'evy-stable and Generalized Hyperbolic laws in Section \ref{sect:background}, we will perform a series of fits of these families to the observed log-returns as well as different statistical tests to quantify their goodness-of-fit (Section \ref{sect:Results}). There, we will also study in close detail the tail behavior of the data in order to elucidate its possible stability and we will address the question of the scaling symmetry of the data. In Section \ref{sect:Discusion} we will sum up the results of the previous section and we will confront them to those obtained in other studies.  The conclusions will be expounded in Section \ref{sect:Conclusions}.

\section{L\'evy-stable laws and \emph{GH} distributions} \label{sect:background}

\subsection{L\'evy-stable laws}

L\'evy-stable laws do not have a closed analytical form for its probability density function in general, but they can be readily defined in terms of their characteristic function $\varphi (t)$:

\begin{equation}
\varphi (t) = \exp [{it\mu - |\delta t|^\alpha(1-i\beta\,\mathrm{sgn}(t)\Phi)}] \\ 
\end{equation}

\begin{equation}
\Phi= \left\{ \begin{array}{rl}
		 \mathrm{tg} \frac{\pi\alpha}{2} & \mbox{if $\alpha \neq 1$} \\
		\\
		-\frac{2}{\pi}\log |t| & \mbox{if $\alpha = 1$}
		\end{array}
	\right.
\end{equation}

The \emph{characteristic exponent} $\alpha \in (0,2]$ determines the weight of the tails and the \emph{skewness parameter} $\beta \in [-1,1]$ its asymmetry. The parameters  $\mu$ and $\delta$ are its \emph{location} and \emph{scale parameters} respectively. The Gaussian distribution is a special case of this family with $\alpha=2, \beta=0$. A random variable $X$ is the limit in distribution of normalized sums of \emph{i.i.d}\ random variables if and only if $X$ has a L\'evy-stable law \cite{breiman1968probability}.

\subsection{Generalized Hiperbolic laws}

The \emph{Generalized Hyperbolic}  distribution can be parametrized in several ways. Following Prause \cite{prause1999generalized}, its probability density function can be written as:

\begin{equation} \label{eqn:GH}
f(x;\lambda, \delta, \alpha,\mu,\beta)=\frac{(\gamma/\delta)^\lambda}{\sqrt{2\pi}K_\lambda(\delta \gamma)} \; e^{\beta (x - \mu)}  \frac{K_{\lambda - 1/2}\left(\alpha \sqrt{\delta^2 + (x - \mu)^2}\right)}{\left(\sqrt{\delta^2 + (x - \mu)^2} / \alpha\right)^{1/2 - \lambda}} 
\end{equation}
where $\gamma=\sqrt{\alpha^2 +\beta^2}$ and $K_{\lambda - 1/2}$ is the modified Bessel function of the third kind with index $\lambda - 1/2$.  The parameter $\alpha >0$ determines the shape of the distribution and $0 \leq |\beta| < \alpha$ its skewness. The usual location and scale parameters are $\mu$ and $\delta$. The parameter $\lambda$ characterizes certain subclasses and influences the size of the mass contained in the tails. These  distributions can be thought as mean-variance mixtures of Gaussian distributions where the mixing distribution is the \emph{Generalized Inverse Gaussian} distribution \cite{eberlein2002generalized}.

Letting $\lambda=-\frac{1}{2}$ we obtain the \emph{Normal Inverse Gaussian} distribution (\emph{NIG}), whose probability density function is: 

\begin{equation}
f(x)= \frac{\alpha\delta K_1 \left(\alpha\sqrt{\delta^2 + (x - \mu)^2}\right)}{\pi \sqrt{\delta^2 + (x - \mu)^2}} e^{\delta \gamma + \beta (x - \mu)}
\end{equation}

All its moments are well defined since it decays as $x^{\alpha}e^{-\beta x}$. The \emph{NIG} distribution is the only subclass of the \emph{GH} family which is closed under convolution; this fact greatly simplifies the computations for option pricing \cite{eberlein2001levy}.

Letting $\lambda = -\frac{\nu}{2}$ and $\alpha \to |\beta|$ in the formula \ref{eqn:GH} above, we obtain the \emph{GH skew Student's t} distribution. Its density is given by:

\begin{equation}
f(x)=\frac{2^{\frac{1-\nu}{2}}\delta^{\nu}|\beta|^{\frac{1+\nu}{2}}K_{\frac{\nu+1}{2}}\left( \sqrt{\beta^{2}(\delta^2 + (x - \mu)^2})\right)e^{\beta (x - \mu)} }{\Gamma (\frac{\nu}{2})\sqrt{\pi}\left(\sqrt{\delta^2 + (x - \mu)^2}\right)^{\frac{\nu+1}{2}}}
\end{equation}

This is the only \emph{GH} subfamily with different asymptotic behaviour of its density function: one tail is a power-law with characteristic exponent equal to $-\nu/2-1$ and the other is a power-law with exponent $-\nu/2-1$ modulated by a factor $e^{-2|\beta||x|}$. If the asymmetry parameter $\beta$ is zero, we recover the classical \emph{Student's t} distribution with symmetric and power-law tails, with a well defined second moment for $\nu>2$ \cite{aas2006generalized}. 

\section{Analysis of the data}
\label{sect:Results}

\subsection{The data}
Our data set contains the price ticks of the index \textsc{Ibex35} of the Madrid Stock Exchange\footnote{Data obtained from \emph{www.tickdata.com}.}. The index \textsc{Ibex35} is a weighted index formed by the 35 most liquid Spanish stocks traded at the Madrid Stock Exchange. The data set covers the period from January 2nd 2009 to December 31st 2010 and comprises 510 market days. 

\begin{figure}[h!]
\centering
    \includegraphics[width=\textwidth]{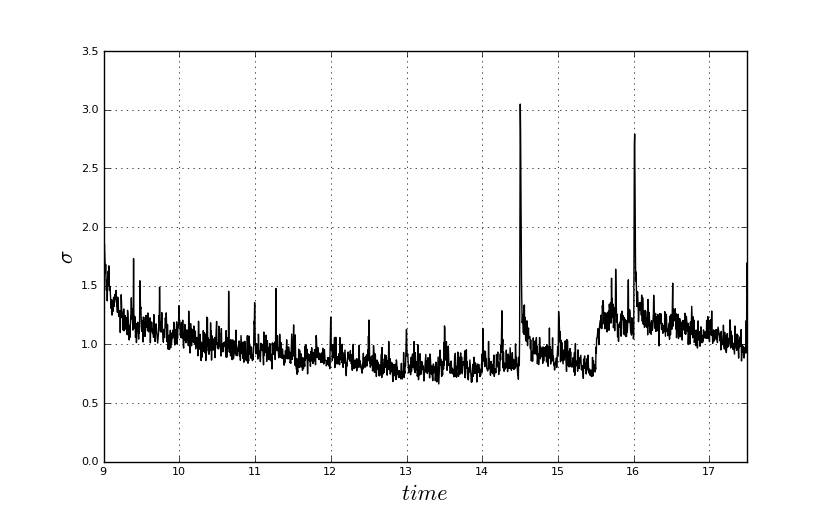}\caption{\textbf{Lunch effect.} At 15:30 CET Wall Street opens, and at 14:30 CET and 16:00 CET macroeconomic indicators in the USA are announced.}
 \label{figure:lunch}

\end{figure}

\begin{figure}[h!]
\centering
    \includegraphics[width=\textwidth]{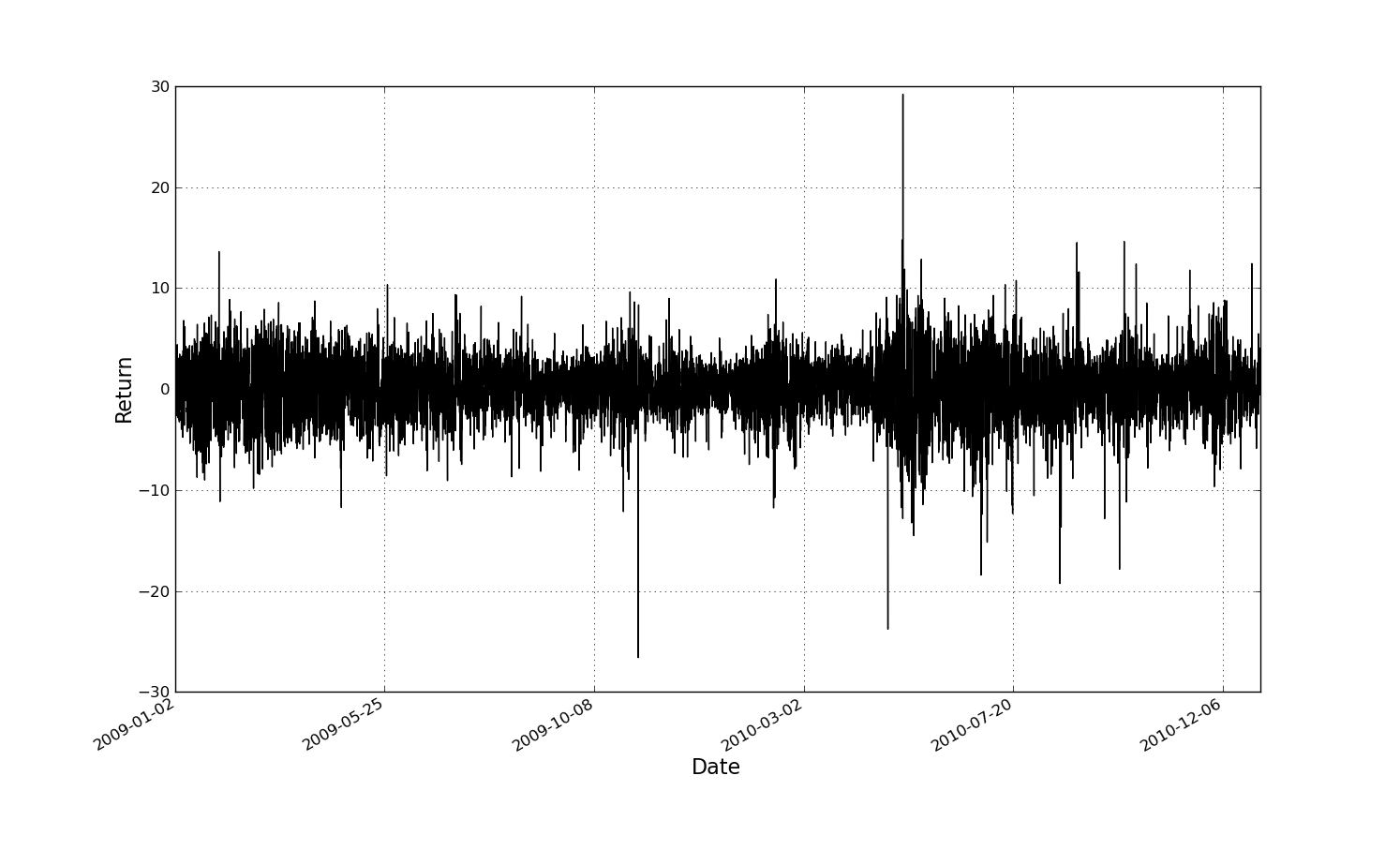}\caption{\textsc{Ibex35} normalized logarithmic returns.}
 \label{figure:returns}

\end{figure}

The values of the index are not updated evenly; the records oscillate between two and around twelve seconds. In order to have a well defined time interval we have sampled these ticks in fifteen-seconds intervals obtaining a series with 1036321 records. From this time series we have obtained the log-returns. However, some issues had to be taken into account before doing this. First,  we have to discard the discontinuity created overnight to avoid artifacts: we therefore focus exclusively on intraday returns. Second,  there is a 30 second uncertainty in the closing time of the session in order to avoid arbitrages: we have accordingly taken a security margin finishing our sessions at 17:29. Some authors \cite{malevergne2005empirical} have also pointed out that the volatility pattern present during the day (the ``lunch effect''; see Figure \ref{figure:lunch}) should be taken into account by normalizing each return  with the average absolute return of that time of the session. However, as happened in the study cited above, in ours we have not observed substantial differences between the raw and the normalized data; therefore, we have worked exclusively with raw returns. The final return series contains 1035810 records, with 2031 records for each market day (Figure  \ref{figure:returns}). The sample statistics can be found in Table \ref{table:samplestatistics} once normalized in scale and location.

\begin{table}
\renewcommand{\arraystretch}{1.2}
\centering
\begin{tabular}{cccccc}
\hline
{\bf MAX.} & {\bf min.} & {$\mathbf{\mu}$} & {$\mathbf{\sigma^2}$} & {$\mathbf{\beta}$} & {$\mathbf{\kappa}$} \\ 
\hline 
29.181 & -28.184 & 0.000 & 1.000 & -0.241 & 13.659 \\
\hline
\\
\end{tabular}
\caption{Sample statistics.}
\label{table:samplestatistics} 
\end{table}

\subsection{Estimation of the parameters}

\begin{figure}[h!]
    \includegraphics[width=\textwidth]{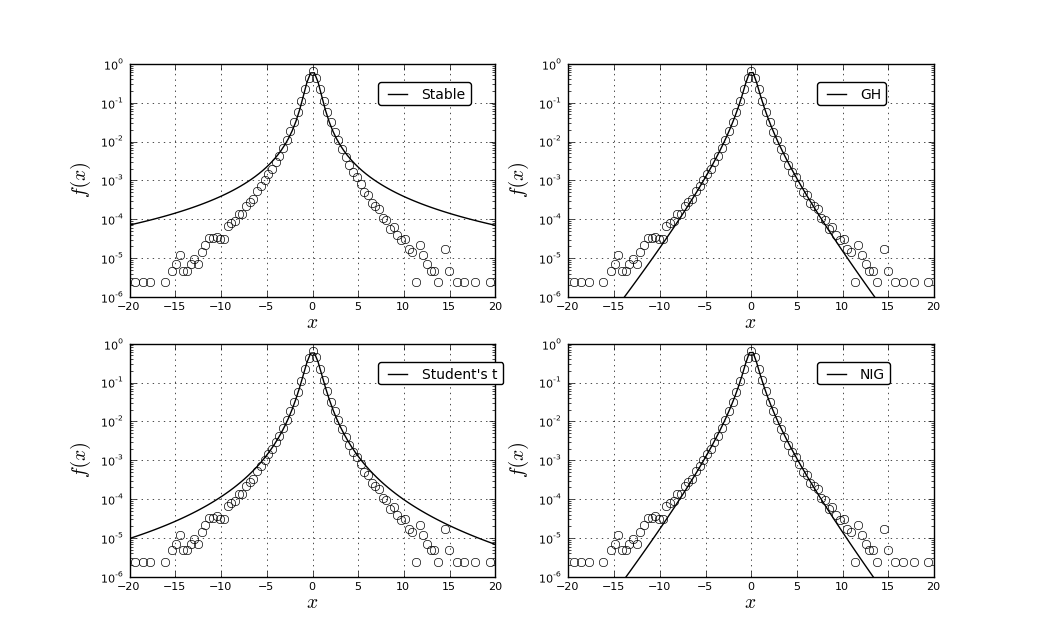}\caption{Histogram and estimated pdfs.}
 \label{figure:histograms}
\end{figure}

Estimation of the parameters of all the distributions has been accomplished using the method of maximum likelihood. The asymptotic properties and optimality of this method of estimation are widely acknowledged \cite{silvey1975statistical}. However,  for the  family of L\'evy-stable laws, parameter estimation via maximum likelihood is not straightforward due to the fact that an analytical expression of the probability density function is not available, and therefore, the method is  only  applicable by numerical approximation which is  very time consuming due to the sample size. Other faster possibilities include methods based on the sample quantiles  \cite{mcculloch1986simple} or regression via the sample characteristic function; see  \cite{weron2004computationally} for a survey of the most usual estimation methods for this family of distributions.  The estimated parameters can be found in Table \ref{table:parameterestimation}, and Figure \ref{figure:histograms} shows  the semilog plots of the estimated densities and the empirical histogram.

\begin{table}[htdp]
\renewcommand{\arraystretch}{1.2}
\centering
\begin{tabular}{lcccccc}
\hline
{Parameters:} &  {\bf  $\mu$} & {\bf $\delta$} & {\bf  $\beta$} & {\bf  $\alpha$} & {\bf $\nu$} & {\bf $\lambda$} \\ 
\hline 
L\'evy-stable &  0.0071 & 0.4825  & 0.0102 & 1.5358 & --- & --- \\
\emph{GH}  & 0.0101  & 0.6495 &  -0.0103 &  0.6296 & --- & -0.5352\\
Student's \emph{t} & 0.0101  & 0.9643 & -0.0089 & --- & 2.7029 & --- \\
\emph{NIG}  & 0.0101  & 0.6365  & -0.0103 & 0.6490  & --- & --- \\
\hline
\\
\end{tabular}
\caption{Estimated distribution parameters.}
\label{table:parameterestimation} 
\end{table}

\subsection{Goodness-of-fit tests}

To quantify the goodness-of-fit of the estimated distributions three different statistical tests have been used: the $\chi ^2$, the Kolmogorov--Smirnov  and the Anderson--Darling  \cite{anderson1954test} tests. These last two tests ---\,based on the cumulative distribution function rather than on the probability density function as the simpler $\chi^2$ test\,--- make a better use of the information contained in the sample since it does not need to be binned. Their drawback, however, is that they  are much more computationally intensive since the distribution function has to be evaluated at the sample points and this implies millions of numerical integrations of trascendental functions. Apart from this, only when the parameters are part of the hypothesis the large-sample distribution of the test statistic is  known; for estimated parameters (like in this case), this distribution is not known except for a few special cases \cite{stephens1976asymptotic}. Montecarlo simulation ---\,the usual approach to tackle this problem \cite{weron2004computationally}\,--- is not feasible here due to the sample size and to the associated computational time needed to obtain the distribution function. Anyway\,---as Anderson \cite{anderson2010anderson} points out---\,the percentage points for the tests when the parameters are estimated are much smaller than those obtained when the parameters are known: a rejected hypothesis using these latter percentage points will then also be rejected with an even higher confidence level when using the former. In any case\,---and as a general rule---\,the lower the value of the test statistic, the better the fit.

% the \emph{p}-values associated with these tests are only reliable when the parameters are given in the hypothesis and not estimated: 

\begin{table}[htdp] 
\renewcommand{\arraystretch}{1.2}
\centering
\begin{tabular}{lccc} 
\hline
{Test statistic}  & {\bf $\chi^2$} & {\bf K-S} & {\bf A-D} \\ 
\hline 

L\'evy-stable  &17133.87 & 0.0175 & 512.17  \\
\emph{GH}  &4556.36 & 0.0147 & 53.88  \\
Student's \emph{t}  & 6967.21 & 0.0148 & 193.73 \\
\emph{NIG}  & 4911.18 & 0.0147  & 52.30  \\
\hline
\\
\end{tabular}
\caption{Goodness-of-fit statistics.}
\label{table:tests} 

\end{table}

\begin{table}[htdp] 
\renewcommand{\arraystretch}{1.2}
\centering
\begin{tabular}{cccc} 
\hline
{\bf Confidence} & {\bf $\chi^2$} & {\bf K-S} & {\bf A-D} \\ 
\hline 
$5\%$ & 231.92 & 0.00134 & 0.4614 \\
$1\%$ & 245.48 & 0.00160 & 0.7435  \\
\hline\\
\end{tabular}
\caption{Critical points for the goodness-of-fit tests.}
\label{table:criticalpoints} 
\end{table}

The values of the test statistics obtained for our data and the upper bounds for the critical points of the tests for the given confidence levels are shown in Tables \ref{table:tests} and \ref{table:criticalpoints}. As it can be observed, the null hypothesis is rejected for any given reasonable confidence level for all the distributions. However, the hyperbolic distributions clearly outperform the L\'evy-stable law. 

For the members of the \emph{GH} family, a likelihood ratio ($\Lambda$)  test has been also performed. This will allow us to quantify which of the two subclasses (\emph{i.e.} \emph{NIG} or Student's \emph{t}) is the soundest and whether or not the \emph{GH} model can be reduced to one of its subfamilies. The values obtained for the statistic $-2\log \Lambda$ are  tabulated in Table \ref{table:likelihoodratio} along with the \emph{p}-values of a $\chi_1 ^2$ variable, its large-sample distribution under the hypothesis of asymptotic normality of the maximum likelihood estimators. According to this, if we accept the hypothesis that the data follows a \emph{GH} distribution, we cannot reject with a confidence level of  $2\%$ the hypothesis that it in fact follows a \emph{NIG} distribution, while the hypothesis that the data follows a Student's \emph{t} distribution is rejected at any reasonable confidence level\footnote{We also performed all the tests for the \emph{Hyperbolic} and \emph{Variance-Gamma} distributions which yielded even worse \emph{p}-values than those obtained for the Student's  \emph{t}. We thus decided not to include them among our results.}.

\begin{table}[htdp] 
\renewcommand{\arraystretch}{1.2}
\centering
\begin{tabular}{ccc} 
\hline
{\bf Distribution} & {\bf $-2\log \Lambda$} & {\bf \emph{p}-value}  \\ 
\hline 
\emph{NIG} & 5.49 & 0.02 \\
\emph{Student's t} & 4551.78 & $< 10^{-16}$  \\
\hline\\
\end{tabular}
\caption{Likelihood-ratio statistics for the \emph{GH} subfamilies}
\label{table:likelihoodratio} 
\end{table}
\subsection{Asymptotic behavior}

\begin{figure}[h!]
  \centering
    \includegraphics[width=\textwidth]{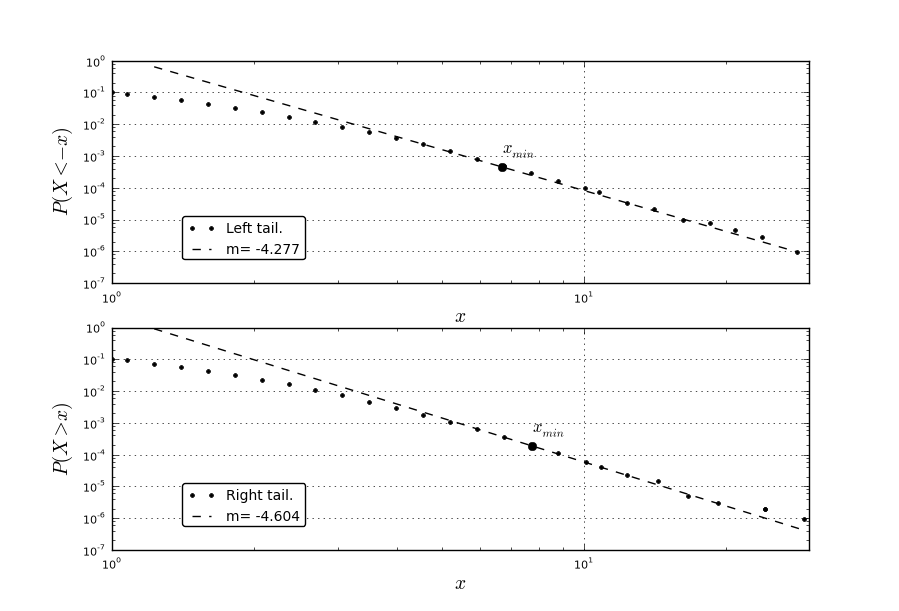}\caption{Tails of the complementary cumulative distribution function.}
 \label{figure:tails}
\end{figure}

Since all of the usual distributions are rejected as plausible hypothesis for the data, we have also studied in detail the asymptotic behavior of the tails. On a log-log plot (Figure \ref{figure:tails}) they seem to fit rather well a straight line. Using the methodology proposed in \cite{clauset2008power} to analyze and estimate power-laws in empirical data, we have obtained a value of 4.60 (resp. 4.28) for the characteristic exponent $\alpha$ and a a value of 7.76 (resp. 6.70) for the scale parameter $x_{min}$ for the right (resp. left) tail.

Even though the tails seem to follow a power-law well outside the stability region, a robust test to reject the hypothesis of stability or even of an exponential behavior is needed. It is well known that for moderate sample sizes an observed tail-index well above two cannot be used as an evidence against stability since it is highly unreliable estimator; if the distribution was really stable, an estimation of the tail parameter using the full sample by maximum likelihood would be more pertinent  \cite{mcculloch1997measuring}.  

\subsection{Stability of the data}

\begin{figure}[h!]
  \centering
    \includegraphics[width=\textwidth]{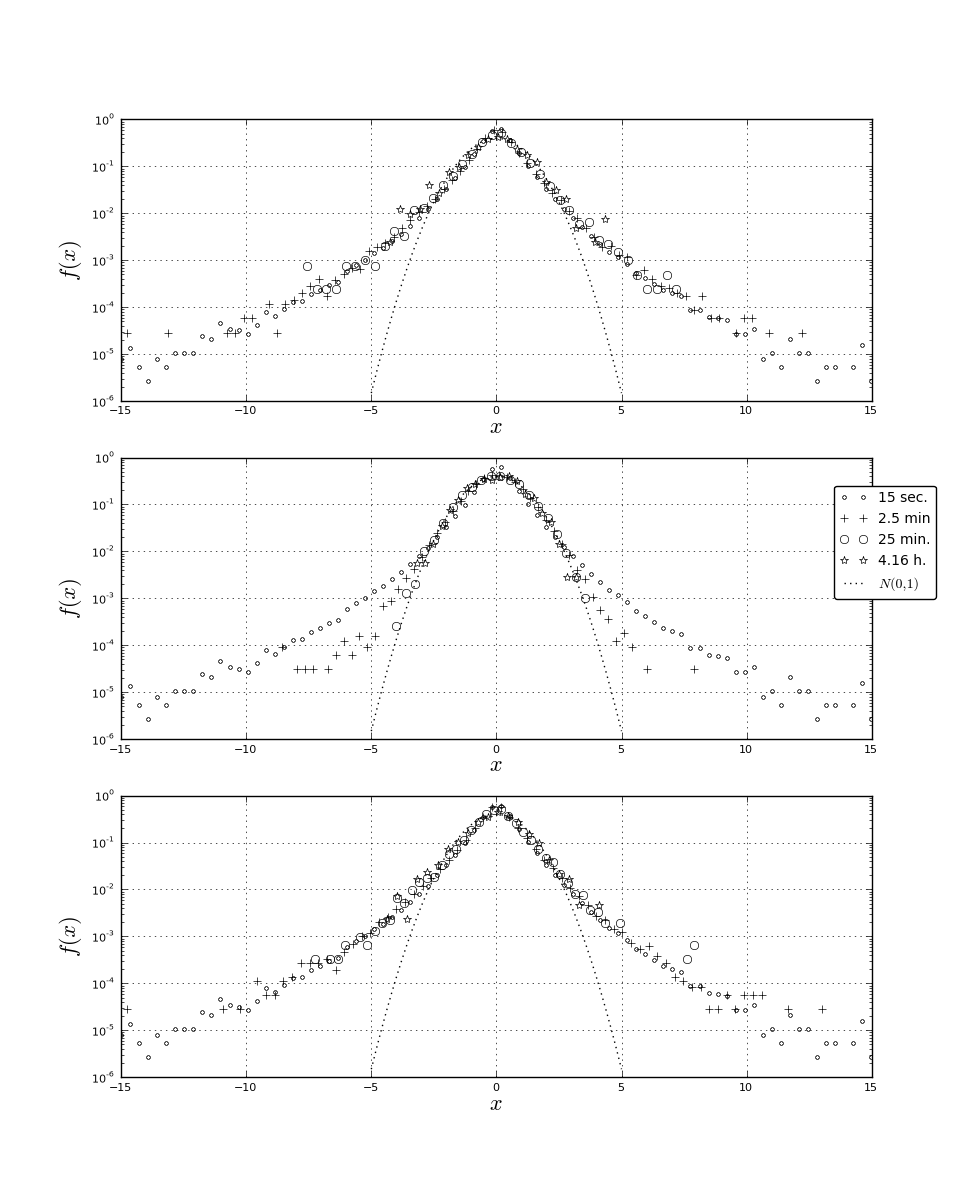}\caption{Rescaled probability distributions of the \textsc{Ibex35} index observed at different time intervals (dotted line: $N(0,1)$). Top: raw data. Middle: reshuffled data. Bottom: daily reshuffled data.}
 \label{figure:scale}
\end{figure}

According to the results of the last paragraph and considering the sample size, the most plausible hypothesis is the lack of stability of the distribution of returns. However, sampling the returns at different time scales $t$ ---\,from fifteen seconds up to half a day\,--- and rescaling it with $t^{1/2}$ its distribution seems to remain stable\footnote{The value of $1/2$ for the scaling exponent was obtained by Detrended Fluctuation Analysis, and it is the one expected for independent identically distributed random variables with finite second moment.}  (Figure \ref{figure:scale}, top panel). This suggests that this symmetry must therefore be the result of the presence of long memory in the data or to the temporal dependence of the parameters.

To support this facts, a reshuffling of the data has been performed (Figure \ref{figure:scale}, middle panel). It can be readily observed that Gaussianity is reached in a few minutes, as could be expected from the central limit theorem. Finally, we have also performed a daily reshuffling of the returns to verify if this scaling symmetry could be an artifact of the \emph{lunch effect}. As can be observed in Figure \ref{figure:scale} (bottom panel), the scaling symmetry of the data still holds, a fact that points towards long--range correlations as the most plausible explanation for this symmetry. It goes without saying that the long memory exhibited by the data and its autocorrelations deserve an in-depth study that shall be addresed in future work.

\section{Discussion}
\label{sect:Discusion}

As far as we know, this is the first study of the high-frequency returns of the \textsc{Ibex35}, so we will necessarily compare our results with those obtained for similar market indexes.

Platen and Sidorowicz, in their study of several world stock indexes \cite{platen2008empirical}, found that for the daily returns of the Madrid stock exchange the best fit was provided by a Student's \emph{t} with 4.51 degrees of freedom, while the \emph{NIG} distribution fit was not as sound. The same results were obtained for a broad group of world stock indexes as an extension of a previous study \cite{hurst1997marginal}. This result is in stark contrast with our findings, where the \emph{NIG} distribution outperforms the other members of the \emph{GH} family as well as the L\'evy-stable laws. As a matter of fact, according to the likelihood-ratio test, the five-parameter \emph{GH} family could be reduced to the four-parameter \emph{NIG} model without much loss.

The L\'evy-stable distributions seem to be also very well suited to model the log-returns of many stock market indexes: this is the case \emph{e.g.}\ of the daily returns of the Hong Kong \textsc{Hang Seng} index. Further, for this index the L\'evy-stable law provides a much better fit than the \emph{NIG} distribution \cite{burnecki2011stability}. Similar results have been observed for the \textsc{IPC} mexican index: the hypothesis of stability could not be rejected at $5\%$ confidence level  while the hypothesis of \emph{NIG} distributed daily log-returns was clearly rejected  \cite{alfonso2012scaling}. However, according to our findings, the L\'evy-stable model is not the best option for modeling the high-frequency returns of the \textsc{Ibex35} since the \emph{NIG} distribution is a much better candidate.  

Considering the tail behaviour of the distribution of log-returns, it is documented that the \textsc{S\&P500} index follows a power law with $\alpha\approx 3$   \cite{gopikrishnan1998inverse, gopikrishnan1999scaling}, while  the characteristic exponent of the  German \textsc{DAX} lies in the range between 3 and 4 \cite{lux2001limiting}.  This is the most commonly accepted range for the characteristic exponent of the tails of the distribution of log-returns. However, the consensus is not complete, and some authors \cite{kinsella2009maximum, malevergne2005empirical} claim that a characteristic exponent in the range  $\alpha \in[3,5]$ could be expected, and even that the decay could well be exponential rather than hyperbolic. We have obtained a seeming power-law behavior for the right (resp.\ left) tail of the distribution with exponent \mbox{$\alpha \approx 4.60$} (resp.\  \mbox{$\alpha \approx 4.28$}), inside the accepted $\alpha \in[3,5]$ range. The only probability distribution analyzed in our study that could have this asymptotic behavior is the Student's \emph{t}. The overall fit, however, rules it out as a plausible model.

The scaling invariance of the financial time series was first proposed and exploited by Mandelbrot in his investigation of the variations of cotton prices \cite{mandelbrot1963variation}. In what regards to stock market indexes, scaling has been observed in the high-frequency returns of \textsc{S\&P500} index  \cite{mantegna1995scaling}  and in the \textsc{OBX} index of the Oslo stock exchange  \cite{skjeltorp2000scaling} among others. In both cases the aggregated log-returns were rescaled with $n^{-1/\alpha}$,  $\alpha$ being the estimated tail parameter for the L\'evy-stable fit  (1.4 for the \textsc{S\&P500} and 1.64 for the \textsc{OBX}, values that are similar to what we have estimated (1.53)). In our study, however, rescaling using the estimated tail exponent for the L\'evy-stable fit  destroys the symmetry;  a value of $1/\alpha=0.5$ obtained by \emph{DFA}\,---and the one expected for finite second moment random variables---\,was used instead.

\section{Summary}
\label{sect:Conclusions}

We have performed a statistical analysis of the high frequency log-returns of the \textsc{Ibex35} index of the Madrid stock exchange over a two year period (2009-2010). Particular attention has been paid to describing the best probability distribution for the data since this question is still controversial in the recent literature, the only fact commonly accepted (although not yet fully incorporated into pricing models) is the departure from normality. Our results show that among the members of the family of Generalized Hyperbolic laws the Normal Inverse Gaussian distribution is the one that provides the best fit for the data. Furthermore, the 5-parameter \emph{GH} family could be well reduced to the 4-parameter \emph{NIG} family without significant loss. This distribution also clearly outperforms the L\'evy-stable laws as a statistical model for this index.

The tails of the distribution of log-returns behave as power laws with exponents $\alpha \approx 4.28$ (left tail) and $\alpha \approx 4.60$ (right tail), a fact that according to the generalized central limit theorem would not be compatible with the stability of the distribution under aggregation. However, the empirical distribution of log-returns has been observed to be stable over several time scales, ranging from a few seconds up to a few hours.  We conjecture that time correlations among the data are probably responsible for this observed stability, since reshuffling the data destroys these time correlations and restores the expected convergence results predicted by the central limit theorem. A more thorough analysis of these time correlations shall be conducted in future work, together with the development of derivative pricing models that take into account more realistic distributions for the underlying assets.

%\bibliographystyle{amsplain}
%\bibliography{finance.bib} 

\end{document}